\documentstyle[11pt,newpasp,twoside,epsf]{article}
\def\edcomment#1{\iffalse\marginpar{\raggedright\sl#1\/}\else\relax\fi}
\reversemarginpar

\newcommand{\be}{\begin{equation}}
\newcommand{\ee}{\end{equation}}
\begin{document}

\title{Using supercluster geometry as a cosmological probe} 
 \author{S. Basilakos}
\affil{Astrophysics Group, Imperial College London, Blackett Laboratory, 
Prince Consort Road, London SW7 2BW, UK}
\author{V. Kolokotronis}
\affil{Institute of Astronomy \& Astrophysics, National Observatory of Athens, 
I.Metaxa \& B.Pavlou, Palaia Penteli, 15236 Athens, Greece}
\author{M. Plionis}
\affil{Institute of Astronomy \& Astrophysics, National Observatory of Athens, 
I.Metaxa \& B.Pavlou, Palaia Penteli, 15326 Athens, Greece}

\begin{abstract}
We study the properties of superclusters detected
in the Abell/ACO cluster catalogue. We identify the superclusters utilizing 
the `friend-of-friend' procedure, and then determine supercluster
shapes by using the differential geometry approach of
Sahni et al. (1998). We find that the dominant supercluster
morphological feature is filamentariness. We compare our Abell/ACO
supercluster results with the corresponding ones generated from two 
different CDM cosmological models in order to investigate statistically 
which of the latter models best reproduces the observational results.
\end{abstract}

\section{Introduction}

The classical pattern of the distribution of matter on
large scales supports the idea that the galaxy clusters
are not randomly distributed but tend to aggregate in
larger systems, the so called superclusters 
(Bahcall 1988 and references therein). 
Individual galaxy superclusters and their properties 
(shape, size etc) have been investigated by different authors (West 1989; 
Jaaniste et al. 1997 and references therein). It has been found that the vast 
majority of the superclusters are flattened triaxial objects, while 
Plionis, Valdarnini, \& Jing (1992) found a preference for prolate 
(filament-like) superclusters. 

In order to study in an objective manner the distribution of superclusters and
their physical properties, it is necessary to develop objective
algorithms and to apply them onto well controlled data.
Indeed different methods like
minimal spanning trees, shape statistics (Sahni \& Coles 1995 references 
therein), genus-percolation statistics (Gout, Melt \& Dickinson 1986) and 
Minkowski functionals (Mecke et al. 1994) have been used in order to describe 
the global geometrical and topological properties of the matter distribution 
utilizing angular and redshift surveys of galaxies as well as N-body 
simulations.

To this end, in this work, we utilize the Abell/ACO cluster catalogue 
in order: (i) to investigate whether we can reliably identify
superclusters and measure their shapes in flux-limited galaxy samples, 
(ii) to measure the shape and size distribution of the Abell/ACO 
superclusters and (iii) to investigate whether these distributions can be 
used as a cosmological probe.

\section{Data and Supercluster Detection}
In our analysis we use the volume-limited 
sample of the Abell/ACO cluster catalogue, with $\mid b \mid \ge 30^{\circ}$ 
and limited within 315$h^{-1}$Mpc or $z\sim 0.11$ (see Einasto et al. 1997). As a result our 
sample contains $\sim 926$ clusters.
In order to find the Abell/ACO superclusters, we use a very common procedure 
based on a `friend-of-friend' algorithm. In particular,  
all mutually linked pairs (within a critical radius)
are joined together to form groups (for those having common boundaries) 
and groups with more than two clusters are identified as candidate 
superclusters. To this end, the critical radius can be defined directly 
from the following equation (Peebles 2001):

\be
R_{\rm cr} \simeq \left(\frac{3-\gamma}{\Omega_{s} \langle n \rangle r_{\circ}^{\gamma}} \right)^{\frac{1}{3-\gamma}}
\ee
where $\Omega_{s}\simeq 6.28$ steradians (solid angle), 
$\langle n \rangle \simeq 1.42 \pm 0.34 \times 10^{-5} h^{3}$Mpc$^{-3}$ (the Abell/ACO mean 
number 
density) and $\gamma=1.8$ with $r_{\circ} \simeq 20 \pm 2 h^{-1}$Mpc 
(the correlation properties). Taking the latter parameters into account, we 
compute $R_{\rm cr}\simeq 27 \pm 4 h^{-1}$Mpc (in agreement with Einasto 
et al. 1997). Finally, using higher or much lower values of $R_{\rm cr}$ we 
tend to connect superclusters and percolate the whole volume. 

\subsection{Shape Statistics}
Shapes are estimated for those ``superclustrers'' that 
consist of 8 or more clusters, utilizing the moments of
inertia ($I_{ij}$) method to fit the best triaxial
ellipsoid to the data (Carter \& Metcalfe 1980). We diagonalize the inertia 
tensor: 
${\rm det}(I_{ij}-\lambda^{2}M_{3})=0$ (where ${\rm M_{3} \;is \; 
3 \times 3 \; unit \; matrix }$), obtaining the eigenvalues $\alpha_{1}$, $\alpha_{2}$, 
$\alpha_{3}$ (where $\alpha_{1}$ is the semi-major axis) 
from which we define the shape of the configuration since, the
eigenvalues are directly related to the three principal axes 
of the fitted ellipsoid. The volume of each supercluster is then 
$V=\frac{4\pi}{3} \alpha_{1} \alpha_{2} \alpha_{3}$.   

The shape statistic procedure, that we use, is based on a differential
geometry approach, introduced by Sahni et al. (1998) 
[for application to the PSCz data see 
Basilakos, Plionis \& Rowan-Robinson 2001]. Here we review only some basic 
concepts. A set of three shapefinders are defined having dimensions of 
length;
${\cal H}_{1}=V S^{-1}$, ${\cal H}_{2}=S C^{-1}$ and ${\cal H}_{3}=C$, 
with $S$ the surface area and $C$ the integrated mean curvature.
Then, it is possible to define a set of two dimensionless shapefinders 
$K_{1}$ and $K_{2}$, as: 
$K_{1}=\frac{ {\cal H}_{2}-{\cal H}_{1} }{ {\cal H}_{2}+{\cal H}_{1} }$
and
$K_{2}=\frac{ {\cal H}_{3}-{\cal H}_{2} } { {\cal H}_{3}+{\cal H}_{2} }$,
normalized to give ${\cal H}_{i}=R$ ($K_{1,2}=0$) for a sphere of radius $R$.
Therefore, based on these shapefinders we can characterize the morphology of
cosmic structures (underdense or overdense regions) 
according to the following categories: (i) pancakes for $K_{1}/K_{2}>1$;
(ii) filaments for $K_{1}/K_{2}<1$; (iii) triaxial for $K_{1}/K_{2}\simeq 1$ and (iv)
spheres for $\alpha_{1} \simeq \alpha_{2} \simeq \alpha_{3}$ and thus $(K_{1},K_{2}) \simeq (0,0)$.
For the quasi-spherical objects the ratio $K_{1}/K_{2}$ 
measures the deviation from pure sphericity.

\section{The Geometrical Properties}
In Figure 1 (left panel) we present the ``shape spectrum'' (broken line) 
and the multiplicity function (open symbols) of the identified
superclusters. From the shape spectrum plot, it is obvious that the dominant feature of the Abell/ACO 
superclusters is filamentariness; ie., $K_{1}/K_{2}<1$ 
(in agreement with previous studies). 
Regarding extreme shaped superclusters, 
we have found, 4 very filamentary superclusters with 
$K_{1}/K_{2} <0.45$ (among which
the Near Shapley and Hercules superclusters),
1 triaxial and 1 extreme pancake-like structures 
with $K_{1}/K_{2}>3$ (one
of which is the Perseus-Pegasus supercluster). 

To complement this, we utilize the completed IRAS flux-limited 
60-$\mu$m redshift survey (PSCz) which is described in
Saunders et al. (2000). We identify, in the
smooth galaxy density field of the PSCz
catalogue, high density regions (superclusters) and
estimate their shapes (for details see Basilakos et al. 2001).
In Figure 1 (right panel) we present a direct comparison 
of the PSCz and Abell/ACO geometrical properties, 
out to 240 $h^{-1}$ Mpc. The two shape-profiles 
are in quite good agreement.
This is a further indication that the two density fields are consistent with 
each other out to this distance.

\subsection{Comparison with Cosmological Models}
We use mock Abell/ACO catalogues 
(having similar to the observed Abell/ACO characteristics)
generated from two 
large cosmological N-body simulations of Colberg et al. (2000), in order 
to investigate whether supercluster properties can discriminate between 
models. In particular, we consider two 
different cold dark matter models covering Hubble volumes, which are:
(1) a flat low-density CDM model with 
$\Omega_{\Lambda}=0.7$ and shape parameter 
$\Gamma=0.17$, $h=0.7$ and $L_{\rm box}=3000 h^{-1}$Mpc and 
(2) a critical density universe with $\Gamma=0.21$ ($\tau$CDM),
$h=0.5$ and $L_{\rm box}=2000 h^{-1}$Mpc.
The CDM models are normalized by the observed cluster abundance at
zero redshift; (Eke, Cole, \& Frenk 1996). We average results over
64 $\Lambda$CDM and 27 $\tau$CDM independent mock Abell/ACO catalogues 
extending out to a radius of 315 $h^{-1}$Mpc. 
We analyse the mock Abell/ACO supercluster properties 
in the same fashion as that of the observed catalogue
and we compare the outcome of this procedure in
Figure 1 (left panel), where we plot the detected supercluster 
shape-spectrum and multiplicity function for two models and Abell/ACO data. 

In order to quantify the
differences between models and data we perform a standard 
$\chi^{2}$ test, assuming Gaussianly distributed errors. This statistical test
proves that the model which is excluded by the data,
by the multiplicity function comparison, 
at a relatively high significance level 
is the $\tau$CDM model (${\cal P}_{>\chi^{2}}=3\times 10^{-3}$), while 
the $\Lambda$CDM model reproduces the 
observed supercluster shape-spectrum 
and multiplicity function (${\cal P}_{>\chi^{2}}=0.80$). 
To validate our analysis we test the robustness of our method by
comparing the models among themselves. We find that the 
shape-spectrum is insensitive to the different cosmologies, 
probably because supercluster shapes reflect
the Gaussian nature of the initial conditions which are common to all
models. However, the supercluster multiplicity function is a strong 
discriminant between the models.

\begin{figure}
\plottwo{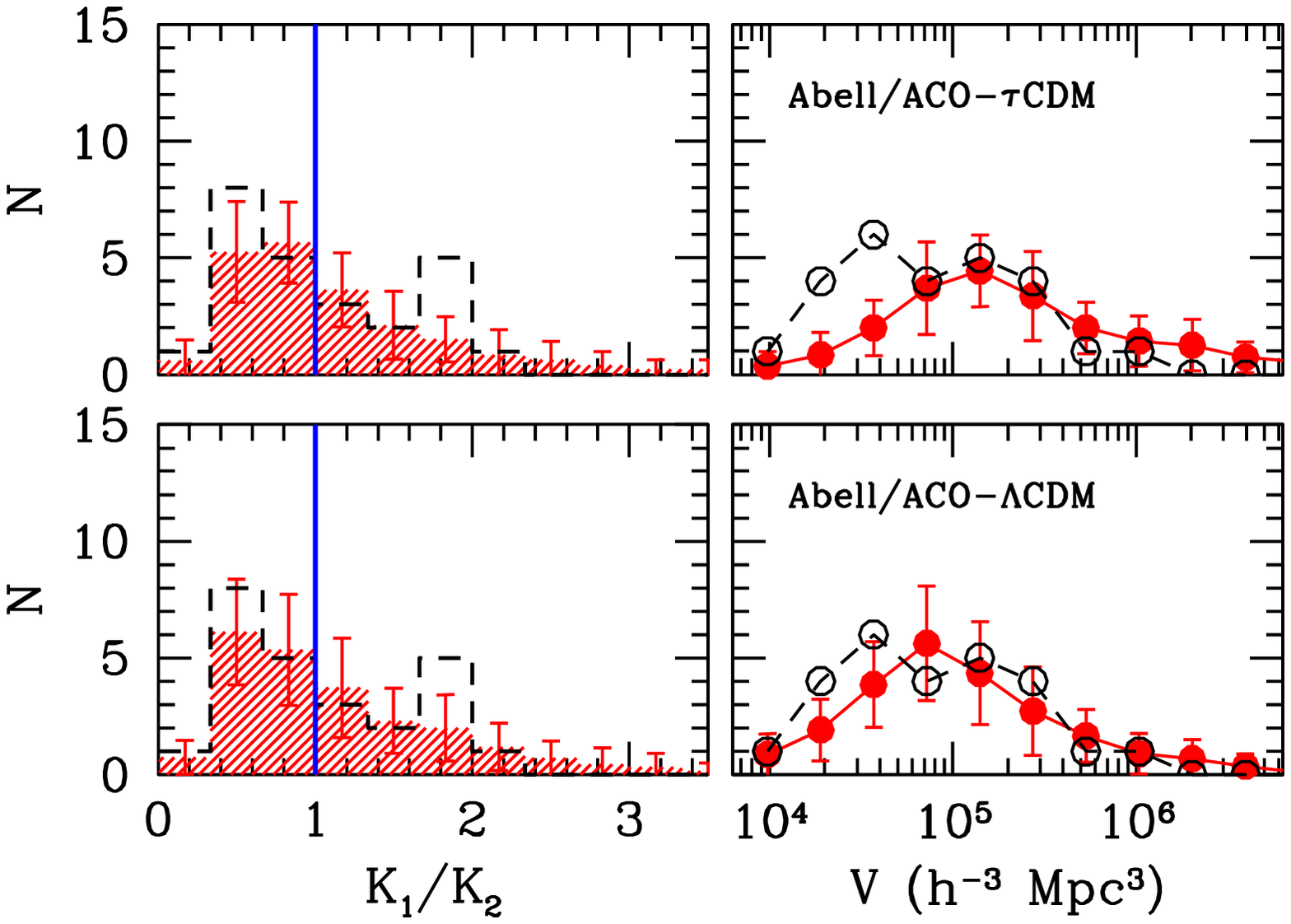}{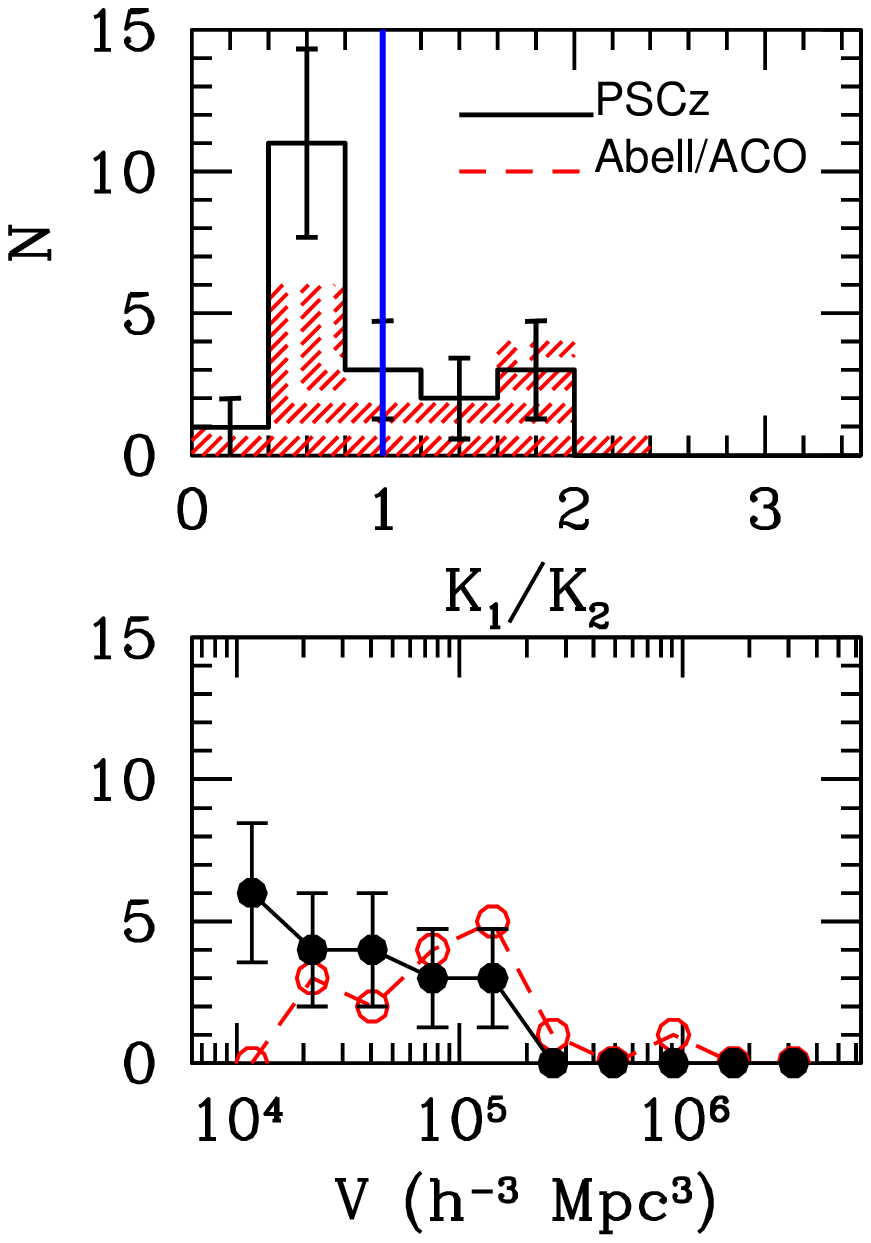}
\caption{{\em Left panel}: 
Comparison with Cosmological models:
The Abell/ACO results are represented by open symbols and
broken lines. {\em Right panel}: Comparison between the Abell/ACO and the PSCz 
geometrical properties. 
}
\end{figure}

\section{Conclusions}
We have studied the properties of superclusters detected in the 
Abell/ACO cluster catalogue.
To determine supercluster shapes we use a differential geometry 
approach and find that the dominant
supercluster morphological feature is filamentariness.
Finally, we have compared our supercluster results with the
corresponding ones generated from the analysis of two cosmological models 
($\tau$CDM and $\Lambda$CDM) and we find that the model that best reproduce 
the observational results is the $\Lambda$CDM model ($\Omega_{\Lambda}=0.7$).


{\small 
\begin{references}
\reference{}Bahcall, N. A., 1988, \araa, 26, 631
\reference{}Basilakos, S., Plionis, M., Rowan-Robinson, M., 2001, \mnras, 223, 47
\reference{}Carter, D. \& Metcalfe, J., 1980, \mnras, 191, 325
\reference{}Colberg, J. M., et al., 2000, \mnras, 319, 209
\reference{}Einasto, M., Tago, E., Jaaniste, J., Einasto, J., Andernach, H.
1997, \aaps, 123, 119
\reference{}Eke, V., Cole, S., Frenk, C. S., 1996, \mnras, 282, 263
\reference{} Gott, J. R., Dickinson, M., Melott, A. L., 1986, \apj, 306, 341
\reference{} Jaaniste, J., Einasto, M., Einasto, J., Andernach, H.,
Muller,  V., 1997, \aap, 329, 1
\reference{} Mecke, K. R., Buchert, T., Wagner, H., 1994, \aap, 288, 697
\reference{}Peebles, P. J. E., 2001, astro-ph/0101127
\reference{}Plionis, M., Valdarnini, R., Jing, Y. P., 1992, \apj, 398, 12
\reference{}Sahni, V. \& Coles, 1995, Phys. REp., 262, 1
\reference{}Sahni, V., Sathyaprokash, B. S., Shandarin, S., 1998a, \apj, 495, L5
\reference{}Saunders, W., et al., 2000, \mnras, 317, 55 
\reference{}West, J. M., 1989, \apj, 347, 610

\end{references}
}
\end{document}